\documentclass[useAMS,usenatbib]{mn2e}
\usepackage{graphicx}
\usepackage{aas_macros}

\def\cyg{Cyg~X-3}
\def\eps{erg~s$^{-1}$}

\title[The relativistic jet of Cyg X-3 in gamma rays]{The relativistic jet of Cygnus X-3 in gamma rays}
\author[G. Dubus, B. Cerutti and G. Henri]{G. Dubus, B. Cerutti and G. Henri\\
Laboratoire d'Astrophysique de Grenoble, UMR 5571 Universit\'e Joseph Fourier Grenoble I / CNRS, BP 53, 38041 Grenoble, France}

\date{Accepted . Received ; in original form \today}

\pagerange{\pageref{firstpage}--\pageref{lastpage}} \pubyear{2010}

\begin{document}

\label{firstpage}

\maketitle

\begin{abstract}
High energy gamma-rays have been detected from \cyg, a system composed of a Wolf-Rayet star and a black hole or neutron star. The gamma-ray emission is linked to the radio emission from the jet launched in the system. The flux is modulated with the 4.8\ hr orbital period, as expected if high energy electrons are upscattering photons emitted by the Wolf-Rayet star to gamma-ray energies. This modulation is computed assuming that high energy electrons are located at some distance along a relativistic jet of arbitrary orientation. Modelling shows that the jet must be inclined and that the gamma ray emitting electrons cannot be located within the system. This is consistent with the idea that the electrons gain energy where the jet is recollimated by the stellar wind pressure and forms a shock. Jet precession should strongly affect the gamma-ray modulation shape at different epochs. The  power in non-thermal electrons represents a small fraction of the Eddington luminosity only if the inclination is low {\em i.e.} if the compact object is a black hole.
\end{abstract}


\section{Introduction}
\cyg\ is a high-mass X-ray binary composed of a compact object in a 4.8\ hr orbit around a Wolf-Rayet (WR) star at a distance of about 7 kpc \citep[see][and references therein]{Bonnet-Bidaud:1988vw,1996A&A...314..521V,2009ApJ...695.1111L}. The system is a bright X-ray source with $L_X\approx 10^{38}$ \eps. \cyg\ is also well-known for radio flaring (up to 20 Jy) when the source has a soft X-ray spectra \citep{2008MNRAS.388.1001S}. 
The radio source is resolved into a relativistic jet with an expansion speed of 0.3-0.7$c $. The strong stellar wind from the WR companion ($\dot{M}_w\approx 10^{-5} M_\odot$yr$^{-1}$, $v_w\approx 1000$ km s$^{-1}$) has a major impact on the environment of the high-energy source. Scattering in the wind is probably responsible for washing out rapid X-ray variability timescales and also for modulating the X-ray emission. It acts as a veil that has made it difficult to identify the nature of the compact object, black hole or neutron star. Despite the differences caused by the WR wind, \cyg\ is firmly established as a trademark accreting binary with relativistic jet {\em i.e.} a microquasar.

The {\em AGILE} and the {\em Fermi Gamma-ray Space Telescope} collaborations have recently reported the detection of high-energy gamma rays (HE, $>$100 MeV) from \cyg\ \citep{2009Natur.462..620T,2009Sci...326.1512F}.  The identification is firm because the detections occur exclusively when \cyg\ is flaring in radio and because {\em Fermi} observations show the HE gamma-ray flux is modulated with the orbital period. The gamma-ray modulation is almost in anti-phase with the X-ray modulation, with the gamma-ray minimum occurring about 0.3-0.4 in phase after X-ray minimum.  The modulation amplitude is  close to $100\%$ after background subtraction. The spectrum is consistent with a power law $F_\nu\sim \nu^{-\alpha}$ with $\alpha=1.7$. The luminosity above 100 MeV is a few $10^{36} (d/{\rm 7\ kpc})^2$ \eps.

Inverse Compton (IC) scattering of photons from the WR star on high energy electrons is a natural candidate to explain the gamma-ray emission. The high temperature of the WR star ($R_\star\approx 1$ R$_\odot$, $T_\star\approx 10^5$\ K) and tight orbit ($d\approx 3\ 10^{11}$ cm) imply that the radiation density in photons from the star is $u_{\star}\approx 10^5$ erg\ cm$^{-3}$ at the location of the compact object, which is at least an order-of-magnitude higher than any other X-ray binary.  Electrons with Lorentz factors of a few 10$^3$ upscatter 20 eV stellar photons above 100 MeV very efficiently in such a radiation field. IC scattering directly produces a modulation of the flux because of the orbital motion. The maximum occurs when stellar photons are backscattered towards the observer.  The accretion disc can also provide seed photons if the HE electrons are close enough. This does not lead to a modulation unless the HE electrons -  disk geometry seen by the observer changes with orbital phase \citep{1977Natur.268..420M}.  
Pion production is possible if there are high energy protons. However, even in this dense environment, it is less efficient than IC so that its energy requirements are higher.

The link between gamma-ray and radio flares suggests that the HE electrons are located in the relativistic jet. Observations of knots in active galactic nuclei show that particles may be accelerated  at specific locations along the jet, linked {\em e.g.} to recollimation shocks \citep{Stawarz:2006oh}. Assuming the electrons mainly upscatter stellar photons at some location along the jet, the expected IC emission will depend upon the distance to the star, the bulk velocity of the jet and its orientation. This orientation is not necessarily perpendicular to the orbital plane if {\em e.g.} the inner accretion disc is warped or it depends on the black hole spin axis. However, the jet orientation is fixed as seen by the observer (changing only if the jet precesses). 

The goal here is to test quantitatively whether the {\em Fermi} gamma-ray modulation can be reproduced in this framework and to see if  constraints can be derived on the jet parameters. 

\section{Jet inverse Compton emission}
\begin{figure}
\centerline{\includegraphics[width=60mm]{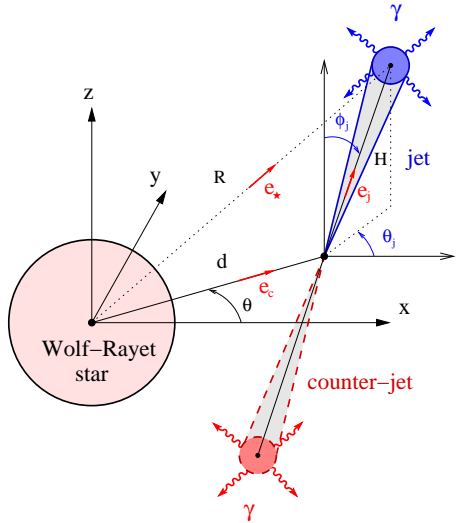}}
\caption{Geometry of the jet model. The scattering electrons are situated at symmetric locations in a jet with relativistic speed $\beta$. The seed photon source is the star.}
\label{geometry}
\end{figure}
\subsection{Emission spectrum}
The HE electrons are assumed to be located at a distance $H$ from the compact object along a jet with a bulk velocity $\beta=v/c$ (Fig.~\ref{geometry}). 
The stellar emission is approximated as a point-like blackbody of temperature $T_\star$ and luminosity $4\pi R_\star^2 \sigma_{SB} T_\star^4$. The electron Lorentz factors $\gamma_e$ are distributed as a power-law $dN_e=K_e \gamma_e^{-p}d\gamma_e$. In the Thompson regime, the inverse Compton emission spectrum at a photon energy $\epsilon$ (in ergs) is given by \citep{Dubus:2010}

\begin{equation}
\begin{array}{ll}
F_{\rm IC}\equiv \epsilon \frac{dN}{dtd\epsilon}=&C(p) K_e  \pi \left(\frac{R_{\star}}{R}\right)^2 \left(kT_{\star}\right)^{\alpha+3}  \\
& \times\ {\cal D}_{\rm obs}^{4+2\alpha}  \left(1-\mathbf{e_\star}.\mathbf{e_{\rm obs}}\right)^{\alpha+1}\epsilon^{-\alpha}
\label{int_bb_end}
\end{array}
\end{equation}
where: the flux index is related to the electron power law index through $\alpha=(p-1)/2$, $R$ is the distance from the star to the electron location; $\mathbf{e_\star}$ and $\mathbf{e_{\rm obs}}$ are unit vectors along, respectively, the star-to-electrons and the electrons-to-observer directions;
\begin{equation}
{\cal D}_{\rm obs}=\frac{(1-\beta^2)^{1/2}}{(1- \beta \mathbf{e_{\rm obs}}.\mathbf{e_{\rm jet}})}
\label{dobs}
\end{equation}
defined the Doppler boost of the jet, $\mathbf{e_{\rm jet}}$ being the unit vector along the jet direction; $C(p)$ is given by 
\begin{equation}
C(p)= \frac{\pi r_e^2 c}{h^3 c^3} \frac{2^{\frac{p+5}{2}}\left(p^2+4p+11\right)\Gamma\left(\frac{p+5}{2}\right)\zeta\left(\frac{p+5}{2}\right)}{\left(p+1\right)\left(p+3\right)\left(p+5\right)}
 \end{equation}
with $\Gamma$ the gamma function and $\zeta$ the Riemann function. This formula is valid in the Thompson regime, that is when $\gamma_e \epsilon_0< m_e c^2$ where $\epsilon_0$ is the characteristic energy of the seed photons. For a blackbody with $T_\star=10^5$ as in \cyg, $\epsilon_0\approx 2.7 kT_\star\approx 23$ eV so the limit occurs for $\gamma_e\approx 2\ 10^4$ (neglecting the Doppler boost). IC emission from 100 MeV to a few GeV (the relevant {\em Fermi} range) occurs in the Thompson regime. 

The model geometry is shown in Fig.~\ref{geometry}. The jet has an azimuth $\theta_{\rm j}$ and polar angle $\phi_{\rm j}$ (=0 when perpendicular to orbital plane). With the origin set at the location of the WR star, 
\begin{equation}
R^2=d^2+H^2 +2  d H(\mathbf{e_{\rm c}}.\mathbf{e_{\rm jet}})
\end{equation}
where $\mathbf{e_{\rm c}}$ is the unit vector along the star to compact object direction, and the unit vectors are given by
\begin{equation}
\begin{array}{lll}
\mathbf{e}_\star&=&(d\mathbf{e}_{\rm c}+H\mathbf{e}_{\rm jet})/R\\
\mathbf{e}_{\rm jet}&=&(\cos\theta_{\rm j} \sin\phi_{\rm j}, \sin\theta_{\rm j} \sin\phi_{\rm j}, \cos \phi_{\rm j})\\
\mathbf{e}_{\rm c}&=&(\cos\theta,\sin \theta, 0)\\
\mathbf{e}_{\rm obs}&=&(0,-\sin i, \cos i)\\
\end{array}
\end{equation}
with  $\theta$ the true anomaly, $d$ the orbital separation and  $i$ the inclination. Here, the true anomaly is defined so that $\theta=\pm\pi/2$ at conjunctions. 

\subsection{Main properties}
The inverse Compton emission has an orbital modulation because of the dependence of $\mathbf{e}_{\rm c}$ on the true anomaly (= orbital phase for a circular orbit). Developing  $\partial F_{\rm IC}/\partial \theta=0$, the emission maximum and minimum along the orbit verify:
\begin{equation}
(\alpha+1)(\mathbf{e_{\rm c}}\times \mathbf{e_{\rm obs}}).\mathbf{e_z}=\frac{H}{R}\left( (\alpha+3)\mathbf{e_\star}.\mathbf{e}_{\rm obs}-2\right) (\mathbf{e_{\rm c}}\times \mathbf{e_{\rm jet}}).\mathbf{e_z}
\end{equation}
If $H\ll d$, or if the jet is perpendicular to the orbital plane, then the maxima and minima are at conjunctions as outlined in \S1. Otherwise, they occur at orbital phases that can be very different. 

The IC flux will be equal to zero if $\mathbf{e}_\star . \mathbf{e}_{\rm obs}=1$ somewhere along the orbit. Having a 100\% modulation can be translated into a necessary condition on $H$ for given $i$, $d$, $\phi_j$ and $\theta_j$. 
Similarly, although the seed photon density decreases with $H$, the  maximum of the IC flux for a given jet geometry does not necessarily occur for $H$=0 because of the dependence of $\mathbf{e}_\star$ on $H$.

The jet speed only appears in $\cal D_{\rm obs}$ and $\mathbf{e_{\rm obs}}.\mathbf{e_{\rm jet}}$ is constant along the orbit: changing $\beta$ will only impact the flux normalisation and not the shape of the modulation. The maximum flux occurs when $\beta=\mathbf{e_{\rm obs}}.\mathbf{e_{\rm jet}}$. Emission from a jet oriented away from the observer will always be weak for highly relativistic speeds because of the deboost.

\begin{figure}
\centerline{\includegraphics{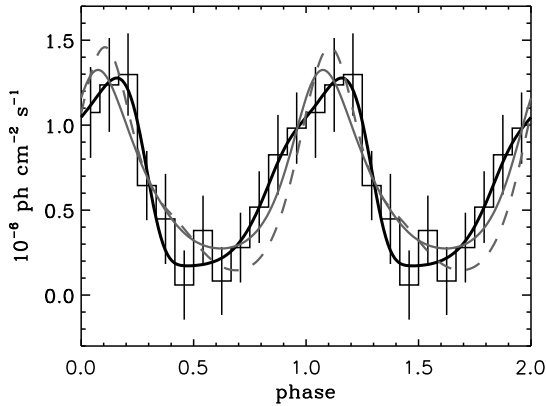}}
\caption{Model fits to the observed $>$ 100 MeV gamma-ray modulation in \cyg. Conjunctions are at phases 0.25 and 0.75 for the conventions adopted in this work. The models shown assumed an orbit with a black hole (O1). The best model is shown with a black solid line. A model with $\beta=0$ is shown with a grey solid line. The model with minimum $P_e$ (3\ $10^{33}$\eps) is shown with a grey dashed lines. All of these models are statistically acceptable fits to the data (see \S3 for details).}
\label{modulation}
\end{figure}
\section{Application to Cyg X-3}
The observed modulation is plotted in Figure~\ref{modulation}. The background level in diffuse gamma rays of 3.6\ 10$^{-6}$ ph\ cm$^{-2}$ s$^{-1}$ was subtracted to the {\em Fermi} lightcurve \citep{2009Sci...326.1512F}. There is not absolute phasing of the orbit of \cyg. The {\em Fermi} observations have been phased so that the well-defined minimum X-ray flux occurs at superior conjunction i.e. phase 0.25 with the conventions adopted in this paper (Fig.~\ref{geometry}). This phasing is justified if the X-ray modulation is due to Thompson scattering in the stellar wind\citep{Pringle:1974oj}. It is independently supported by infrared spectroscopy \citep{2000ApJ...541..308H}.

The orbital parameters of \cyg\ are not determined precisely \citep{2000ApJ...541..308H, 2009A&A...501..679V} so two extreme solutions are adopted following \citet{2008MNRAS.386..593S}. Orbit 1 (O1) has a $M_1$=20 M$_\odot$ black hole around a 50 M$_\odot$ WR star of radius 2.3 $R_\odot$ and is seen with an inclination of 30$\degr$. Orbit 2 (O2) has a $M_1$=1.4 M$_\odot$ neutron star around a 5 M$_\odot$ WR star of radius 0.6 $R_\odot$ with $i=70\degr$. The {\em Fermi} spectrum $\alpha=1.7$ sets the electron power-law index $p=4.4$. The emission arise from two symmetric sites: the jet and the counterjet. The counterjet has $\phi_{cj}=\pi+\phi_j$.

\subsection{Parameter exploration}
The jet is parametrised by $\beta$, $H$, $\phi_j$, $\theta_j$ and $K_e$. The expected modulation in the {\em Fermi} band is calculated using the equation in \S2\ for the jet and the counterjet. The evaluation of Eq.~\ref{int_bb_end} is very fast and allows an exhaustive exploration of the parameter space. The jet angle $\phi_j$ was varied between 0 and $\pi$/2 ; $\theta_j$ varied between 0 and 2$\pi$. The emission height $H$ was varied between 0.01$d$ and 100$d$ in logarithmic steps ($d$ is the orbital separation). The jet speed $\beta$ was varied linearly from 0 to 0.99 (bulk Lorentz factor $\approx 7$).  

The model $K_e$ is adjusted to minimize the $\chi^2$ goodness-of-fit to the observed modulation. The normalisation $K_e$ is converted into a power in HE electrons $P_e$  assuming a distance of 7 kpc and a minimum HE electron Lorentz factor $\gamma_{\rm e, min}=1000$. $P_e$ is highly sensitive to $\gamma_{\rm e, min}$ because of the very steep electron spectrum. IC emission above 100 MeV requires that $\gamma_{\rm e, min}\leq 1000$ so $P_e$ is a lower limit on the non-thermal power. 

Good fits can be obtained for both O1 ($\chi^2_{\rm min}=2.7$ for 12 data points - 5 parameters = 7 degrees of freedom) and O2 ($\chi^2_{\rm min}=4.2$). The best model for O1 is plotted in Figure~\ref{modulation}. It has $\beta=0.41$, $H=8\ 10^{11}$cm, $\phi_j=39\degr$, $\theta_j=319\degr$, $P_e=10^{38}$\eps. The 90\% confidence range for the parameters was determined by adding 9.24 to the minimum $\chi^2$ \citep{1976ApJ...208..177L}.  Only models that had $P_e$ lower than the Eddington luminosity $L_{\rm Edd}\approx\ 10^{38} (M_1/M_\odot)$ erg\ s$^{-1}$ were kept. Besides being physically implausible, models with larger $P_e$ are associated with high values of $\beta$ or large $H$. The high $P_e$ then compensates for Doppler deboosting or low IC efficiency (see \S3.3).

\subsection{Jet orientation}
\begin{figure}
\centerline{\includegraphics{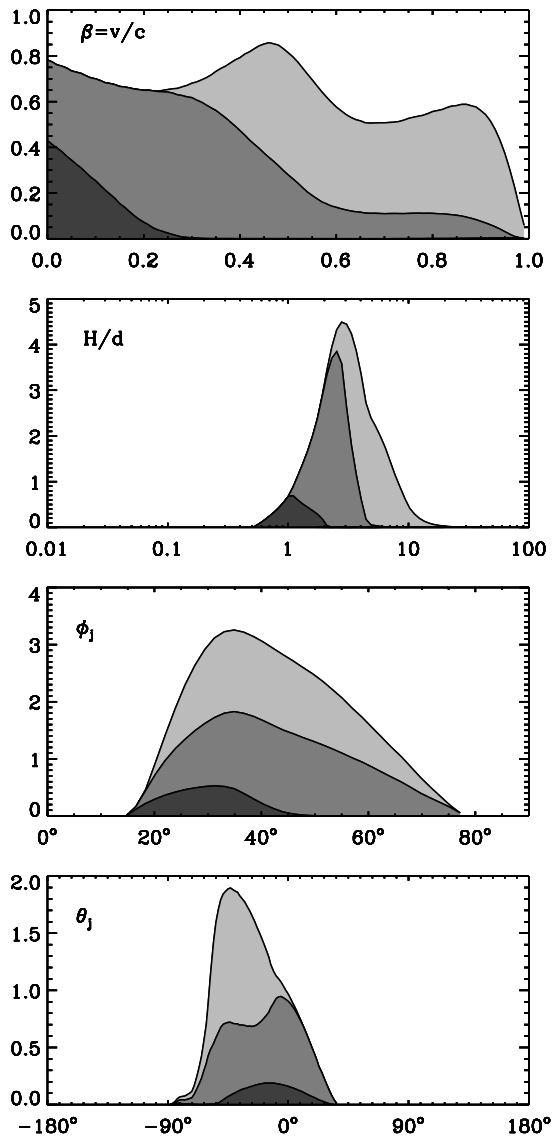}}
\caption{Distribution of jet parameters for models in the 90\% confidence region given by $\chi^2$ statistics. Orbit O1 (20 M$_\odot$ black hole, $i$=30$\degr$) is assumed. The various regions correspond to a power in high energy electrons $P_e\leq L_{\rm Edd}$ (light grey), $\leq$ 0.1$L_{\rm Edd}$ (grey), $\leq$ 0.01$L_{\rm Edd}$ (dark grey). Here, $L_{\rm Edd}$ is 2\ 10$^{39}$ \eps.}
\label{distri}
\end{figure}
Figure~\ref{distri} shows the distributions of $\beta$, $H$, $\phi_j$ and $\theta_j$ for the black hole case (O1). The figure also shows the distributions for various limits on $P_e$. In all cases, the HE electrons distance $H$ is between 0.5 and 30 times the orbital separation (i.e. between 2\ 10$^{11}$ and 10$^{13}$ cm). A location very close to the compact object is excluded. The orientation of the jet is constrained to be $20\degr \la \phi_j\la 80\degr$ with a preference for values comparable to the system inclination ($i=30\degr$). A jet perpendicular to the orbital plane does not fit the data. The azimuth $\theta_j$ is less constrained: there is a well defined peak in the distribution (bottom panel, Fig.~\ref{distri}) but, contrary to $H$ or $\phi_j$,  there are good models all over the range even if in small numbers (not visible on a linear scale).

Moderate relativistic speeds $\beta$ are favoured but this is not strongly constrained. The speed is closely linked to the power in HE electrons.  There is a tendency to have lower values of $\beta$ when the allowed $P_e$ gets smaller, accompanied by a smaller $H$. A model in the 90\% confidence region with $\beta$=0 is shown in Figure~\ref{modulation}. It has $\chi^2=7.1$, $H=7\ 10^{11}$cm, $\phi_j=31\degr$, $\theta_j=9\degr$, $P_e=2\ 10^{37}$\eps.  This trend on $\beta$ reverses for low values of $P_e \la$ 0.001 $L_{\rm Edd}$. These do not appear  in Figure~\ref{distri} as there are comparatively very few such models.  The minimum $P_e$ in the 90\% confidence region is 4\ 10$^{33}$ \eps, a very modest fraction of $L_{\rm Edd}$. This model is also shown in Figure~\ref{modulation}. It has  $\chi^2=11.3$, $\beta=0.99$, $H=10^{12}$cm, $\phi_j=32\degr$ and $\theta_j=275\degr$. These low $P_e$ models all have $\phi_j\approx i$ and $\theta_j\approx =-90\degr$: they are almost aligned with the observer ($\mathbf{e_{\rm jet}}. \mathbf{e_{\rm obs}}\approx 1$) at superior conjunction. The slight difference in  $\theta_j$ accounts for the phase difference of the maximum. Here, Doppler boosting compensates for the low $P_e$. There is some degeneracy between the two parameters up to some (large) value of the Lorentz factor $\approx 20$ where good models cannot be found anymore. These are effectively microblazar models.

The constraints in the neutron star case (orbit O2, not shown here) are similar. The jet orientation is well constrained with  $25\degr \la \phi_j\la 65\degr$,  $-60\degr \la \theta_j\la -10\degr$ and $2\ 10^{11} {\rm \ cm}\la H\la 6\ 10^{11}{\rm \ cm}$ ($H/d$ from 1 to 3), comparable to the values found with O1. However, in all cases $\beta$ is $\la$ 0.2. Interestingly, $P_e$ is constrained to be rather large with $P_e \ga 0.2 L_{\rm Edd}$ (about $3\ 10^{37}$\eps). The large inclination (70$\degr$) required for a neutron star primary is the reason for the difference with the black hole case. Arbitrarily setting $i=30\degr$ with the orbit O2 gives results for $\beta$ and $P_e$ that are consistent with those of O1. Large inclinations do not allow good fits for small values of $P_e$ or large values of $\beta$.

These results were obtained for a steep power-law distribution of electrons with an index $p=4.4$, because of the soft gamma-ray flux index and the assumption of Thompson scattering. Taking  $p=2$ or $p=3$ does not affect the conclusions. A few tests calculations using the full IC cross section (done as explained in \citealt{Dubus:2010}) showed that a slightly harder electron index ($p\approx 4$) is required to match the spectrum. Again,  this does not change the results. The steep spectrum may not directly reflect an electron power-law distribution but represent the best fit to {\em e.g.} a cutoff in the 100 MeV -- 1 GeV range. To test this, a lightcurve was calculated (including the full IC cross section) for a jet with the parameters of the best fit shown in Fig.~\ref{modulation} but assuming a power law distribution $p=3$ from $\gamma_{\rm e}=100$ up to $\gamma_{\rm e, cutoff}\approx 3\ 10^3$. (A $p=3$ slope is expected for a steady state distribution of electrons injected with the canonical $p=2$ power law in the presence of strong Thompson IC cooling.) The $>$100 MeV lightcurve was indistinguishable from the one in Fig.~\ref{modulation}, even though the cutoff energy changed significantly along the orbit due to Doppler boosting. Hence, the results obtained here are likely to extend when more complex spectral shapes and Klein-Nishina effects are taken into account.

\subsection{Jet precession}
\begin{figure}
\centerline{\includegraphics{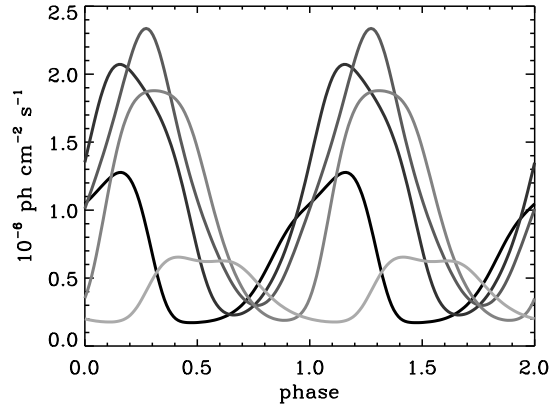}}
\caption{Impact of jet precession on the gamma-ray lightcurve for the best-fit model shown in Figure~\ref{modulation}. The jet azimuth $\theta_j$  is rotated in steps of 72$\degr$ from its best fit value of 319$\degr$, with lighter lines as $\theta_j$ moves away from this value.}
\label{precession}
\end{figure}
The preceding section showed that the jet must be inclined in order to obtain good fits to the gamma-ray modulation. There is evidence for jet inclination in \cyg\ as well as other microquasars \citep{Maccarone:2002so}. An inclined jet is likely to undergo precession on a timescale longer than the orbital period. There is currently no evidence for or against jet precession in \cyg. 
Here, jet precession will manifest itself as a change in the gamma-ray modulation since  $\theta_j$ will sample the full range from 0 to $2\pi$ in a full precession. Both the shape and amplitude are affected as shown in Figure~\ref{precession}. The peak flux phase and amplitude can vary dramatically from one precession phase to another. 

The {\em Fermi} data already show a hint for a change in the phasing of the modulation between the two epochs during which \cyg\ was detected. In addition, the first reported detection of \cyg\ at 100 MeV from SAS-2 showed a gamma-ray orbital modulation correlated (instead of roughly anti-correlated) with the X-ray modulation \citep{Lamb:1977xw}. Later observations by {\em Cos B} and EGRET failed to re-detect the source unambiguously \citep{Mori:1997nu}. A possible explanation is that the jet orientation had changed in between these observations. Future {\em Fermi} observations of \cyg\ may find a different modulation lightcurve or may actually fail to detect the source because of its low flux, even though \cyg\ shows the right radio and X-ray state. 

The comparison between gamma-ray lightcurves can serve as a very powerful diagnostic of the jet geometry. For instance, in the microblazar models discussed in \S3.2, the near perfect alignment of a jet with the line-of-sight and the high $\beta$ means that the gamma-ray flux is detectable only during the very short interval in precession phase where it is Doppler boosted. The gamma-ray flux will be deboosted most of the time --- so that the {\em Fermi} and {\em AGILE} detections  would have required very special circumstances.

\section{Conclusions}
The orbital modulation of the $>$100 MeV flux from \cyg\ can be very well fitted by a simple-minded model in which the emission is due to HE electrons up-scattering stellar photons. The HE electrons are situated in two symmetric locations in a relativistic jet with an arbitrary orientation. 

The fitting procedure reveals that the jet is necessarily inclined to the orbital plane normal. The most likely value is close to the line-of-sight ($\phi_j\approx i$, in agreement with the conclusions based on radio imaging of the jet \citep{2001ApJ...553..766M}.  The HE electrons cannot be close to the compact object. They are outside of the system at distances of at least one orbital separation, possibly up to 10$d$. IC scattering of accretion disc photons is then irrelevant. If the compact object in \cyg\ is a neutron star, the required power in HE electrons is a significant fraction of the Eddington luminosity. For a black hole, because of the lower system inclination implied, the power required can be as low as $10^{-5} L_{\rm Edd}$. These conclusions appear robust even when more complex electron distributions and the full IC cross-section are taken into account. Precession can be expected from an inclined jet. It should cause a change in the shape and amplitude of the gamma-ray modulation in the future.

The IC cooling timescale is $t_{\rm ic} \approx 0.5 (\gamma_e/10^3)^{-1} (R/d)^2$ seconds (scaled to the orbital separation $d$ and for orbit O1). The size of the gamma ray emitting region is roughly $s\approx \beta c t_{\rm ic}$, giving $s/R\la 0.04 \beta(\gamma_e/10^3)^{-1} (R/d)$ when scaled to $R$. Hence, the assumption that the emission in the {\em Fermi} energy range is localised holds up to distances $\approx 10d$ from the star. Cooling slows down at lower energies and electrons emit synchrotron radio beyond the $\gamma$-ray emission zone on much larger scales. 

The $\gamma$-ray emission zone could be related to electron acceleration at a recollimation shock as the jet pushes its way through the stellar wind. The jet is initially over-pressured compared to its environment. It expands freely until its pressure $p_j$ matches that of the environment $p_e$. Here, $p_e$ is the ram pressure of the supersonic wind $\rho_w v_w^2$. The jet pressure is $p_j\sim L_j/(\pi c \Theta^2 l^2)$ where $L_j$ is the jet power, $\Theta$ is its opening angle and $l$ is the distance along the jet \citep[e.g.][]{1997MNRAS.287L...9B}. The pressures equilibrate at
\begin{equation}
\frac{l}{R}\sim 0.5\ \Theta^{-1} {L}_{38}^{1/2} \dot{M}_{-5}^{-1/2} {v}_{1000}^{-1/2} 
\end{equation}
with $L_j=10^{38}\ {\rm erg\ s}^{-1}$, $\dot{M}_{w}=10^{-5} M_\odot\ {\rm yr}^{-1}$ and ${v}_{w}=1000\ {\rm km\ s}^{-1}$. A jet recollimation shock forms beyond $l$. The shock crosses the jet axis after a further distance of order $l$ when the external pressure is constant \citep{Stawarz:2006oh}. This is roughly the case here since the jet does not extend very far from the system and the dependence of $p_w$ with $l$ remain shallow (unless it is pointed directly away from the star). The location is consistent with the values of $H$ derived above, suggesting this is where jet kinetic or magnetic energy is channeled into particle acceleration. This should be verified by calculations taking into account the non-radial nature of the jet-wind interaction. The shock occurs in the wind only because $\dot{M}_{w}$ is very large (WR star) and the orbit very tight. Most microquasar jets will actually break out of the immediate vicinity of the system and interact much further away when their pressure matches that of the ISM. Any HE particles there will find a much weaker radiation environment and will be less likely to produce a (modulated) IC gamma-ray flux detectable by {\em Fermi} or {\em AGILE}.

The emerging picture is that of a jet launched around a black hole, with a moderate bulk relativistic speed, oriented not too far from the line-of-sight, interacting with the WR stellar wind to produce a shock at a distance of 1-10$d$ from the system, where electrons are accelerated to GeV energies and upscatter star photons.
\section*{Acknowledgments}
The authors thank S. Corbel, J.-P. Lasota, L. Stawarz and A. Szostek for comments.  This work was supported by the European Community via contract ERC-StG-200911.

\bibliographystyle{mn2e}
\bibliography{../BIBLIO}
\end{document}